\documentclass[10pt,conference]{IEEEtran}
\IEEEoverridecommandlockouts
\usepackage{cite}
\usepackage{amsmath,amssymb,amsfonts}
\usepackage{graphicx}
\usepackage{textcomp}
\usepackage{xcolor}
\usepackage{hyperref}
\usepackage{multirow}
\usepackage{algorithm}
\usepackage{algpseudocode}
\usepackage{units}
\usepackage[nolist,nohyperlinks]{acronym}
\usepackage{bm}

\usepackage{tikz}
\usetikzlibrary{%
    angles,%
    arrows,%
    arrows.meta,%
    babel,%
    backgrounds,%
    calc,%
    fit,%
    matrix,%
    mindmap,%
    patterns,%
    positioning,%
    quotes,%
    shapes,%
    shadows,%
    tikzmark,%
    trees,%
}
\usepackage{pgfplots}
\usepgfplotslibrary{groupplots}
\usepgfplotslibrary{statistics}
\pgfplotsset{width=\textwidth*0.4,compat=1.9}

\usepackage{xcolor}

\acrodef{DT}[DT]{Digital twin}
\acrodef{RAN}[RAN]{radio access network}
\acrodef{GAN}[GAN]{generative adversarial network}
\acrodef{GMM}[GMM]{Gaussian mixture model}
\acrodef{HMM}[HMM]{hidden Markov model}
\acrodef{IAT}[IAT]{inter arrival time}
\acrodef{MDN}[MDN]{mixture density network}
\acrodef{EM}[EM]{expectation-maximization}
\acrodef{HTTP}[HTTP]{hypertext transfer protocol}
\acrodef{UDP}[UDP]{datagram protocol}
\acrodef{MTU}[MTU]{Ethernet maximum transmission unit}
\acrodef{NLL}[NLL]{negative log likelihood}
\acrodef{GRU}[GRU]{gated recurrent unit}
\acrodef{CDF}[CDF]{cumulative distribution functions}
\acrodef{KL}[KL]{Kullback–Leibler}
\acrodef{KS}[KS]{Kolmogorov-Smirnov}
\acrodef{PSD}[PSD]{power spectral density}
\acrodef{ACF}[ACF]{autocorrelation function}

\def\BibTeX{{\rm B\kern-.05em{\sc i\kern-.025em b}\kern-.08em
    T\kern-.1667em\lower.7ex\hbox{E}\kern-.125emX}}

\def\checkmark{\tikz\fill[scale=0.4](0,.35) -- (.25,0) -- (1,.7) -- (.25,.15) -- cycle;} 

\usepackage{tikz}
\usetikzlibrary{%
    angles,%
    arrows,%
    arrows.meta,%
    babel,%
    backgrounds,%
    calc,%
    fit,%
    matrix,%
    mindmap,%
    patterns,%
    positioning,%
    quotes,%
    shapes,%
    shadows,%
    tikzmark,%
    trees,%
}
\usepackage{pgfplots}
\usepgfplotslibrary{groupplots}
\usepgfplotslibrary{statistics}
\pgfplotsset{width=\textwidth*0.4,compat=1.9}

\begin{document}

\definecolor{kit-green}{RGB}{0, 150, 130}
\colorlet{KITgreen}{kit-green}
\definecolor{kit-green100}{RGB}{0, 150, 130}
\definecolor{kit-green90}{rgb}{0.1, 0.6294, 0.5588}
\definecolor{kit-green80}{rgb}{0.2, 0.6706, 0.6078}
\definecolor{kit-green75}{rgb}{0.25, 0.6912, 0.6324}
\definecolor{kit-green70}{rgb}{0.3, 0.7118, 0.6569}
\definecolor{kit-green60}{rgb}{0.4, 0.7529, 0.7059}
\definecolor{kit-green50}{rgb}{0.5, 0.7941, 0.7549}
\definecolor{kit-green40}{rgb}{0.6, 0.8353, 0.8039}
\definecolor{kit-green30}{rgb}{0.7, 0.8765, 0.8529}
\definecolor{kit-green25}{rgb}{0.75, 0.8971, 0.8775}
\definecolor{kit-green20}{rgb}{0.8, 0.9176, 0.902}
\definecolor{kit-green15}{rgb}{0.85, 0.9382, 0.9265}
\definecolor{kit-green10}{rgb}{0.9, 0.9588, 0.951}
\definecolor{kit-green5}{rgb}{0.95, 0.9794, 0.9755}

\definecolor{kit-blue}{RGB}{70, 100, 170}
\colorlet{KITblue}{kit-blue}
\definecolor{kit-blue100}{RGB}{70, 100, 170}
\definecolor{kit-blue90}{rgb}{0.3471, 0.4529, 0.7}
\definecolor{kit-blue80}{rgb}{0.4196, 0.5137, 0.7333}
\definecolor{kit-blue75}{rgb}{0.4559, 0.5441, 0.75}
\definecolor{kit-blue70}{rgb}{0.4922, 0.5745, 0.7667}
\definecolor{kit-blue60}{rgb}{0.5647, 0.6353, 0.8}
\definecolor{kit-blue50}{rgb}{0.6373, 0.6961, 0.8333}
\definecolor{kit-blue40}{rgb}{0.7098, 0.7569, 0.8667}
\definecolor{kit-blue30}{rgb}{0.7824, 0.8176, 0.9}
\definecolor{kit-blue25}{rgb}{0.8186, 0.848, 0.9167}
\definecolor{kit-blue20}{rgb}{0.8549, 0.8784, 0.9333}
\definecolor{kit-blue15}{rgb}{0.8912, 0.9088, 0.95}
\definecolor{kit-blue10}{rgb}{0.9275, 0.9392, 0.9667}
\definecolor{kit-blue5}{rgb}{0.9637, 0.9696, 0.9833}

\definecolor{kit-red}{RGB}{162, 34, 35}
\colorlet{KITred}{kit-red}
\definecolor{kit-red100}{RGB}{162, 34, 35}
\definecolor{kit-red90}{rgb}{0.6718, 0.22, 0.2235}
\definecolor{kit-red80}{rgb}{0.7082, 0.3067, 0.3098}
\definecolor{kit-red75}{rgb}{0.7265, 0.35, 0.3529}
\definecolor{kit-red70}{rgb}{0.7447, 0.3933, 0.3961}
\definecolor{kit-red60}{rgb}{0.7812, 0.48, 0.4824}
\definecolor{kit-red50}{rgb}{0.8176, 0.5667, 0.5686}
\definecolor{kit-red40}{rgb}{0.8541, 0.6533, 0.6549}
\definecolor{kit-red30}{rgb}{0.8906, 0.74, 0.7412}
\definecolor{kit-red25}{rgb}{0.9088, 0.7833, 0.7843}
\definecolor{kit-red20}{rgb}{0.9271, 0.8267, 0.8275}
\definecolor{kit-red15}{rgb}{0.9453, 0.87, 0.8706}
\definecolor{kit-red10}{rgb}{0.9635, 0.9133, 0.9137}
\definecolor{kit-red5}{rgb}{0.9818, 0.9567, 0.9569}

\definecolor{kit-yellow}{RGB}{252, 229, 0}
\colorlet{KITyellow}{kit-yellow}
\definecolor{kit-yellow100}{RGB}{252, 229, 0}
\definecolor{kit-yellow90}{rgb}{0.9894, 0.9082, 0.1}
\definecolor{kit-yellow80}{rgb}{0.9906, 0.9184, 0.2}
\definecolor{kit-yellow75}{rgb}{0.9912, 0.9235, 0.25}
\definecolor{kit-yellow70}{rgb}{0.9918, 0.9286, 0.3}
\definecolor{kit-yellow60}{rgb}{0.9929, 0.9388, 0.4}
\definecolor{kit-yellow50}{rgb}{0.9941, 0.949, 0.5}
\definecolor{kit-yellow40}{rgb}{0.9953, 0.9592, 0.6}
\definecolor{kit-yellow30}{rgb}{0.9965, 0.9694, 0.7}
\definecolor{kit-yellow25}{rgb}{0.9971, 0.9745, 0.75}
\definecolor{kit-yellow20}{rgb}{0.9976, 0.9796, 0.8}
\definecolor{kit-yellow15}{rgb}{0.9982, 0.9847, 0.85}
\definecolor{kit-yellow10}{rgb}{0.9988, 0.9898, 0.9}
\definecolor{kit-yellow5}{rgb}{0.9994, 0.9949, 0.95}

\definecolor{kit-orange}{RGB}{223, 155, 27}
\colorlet{KITorange}{kit-orange}
\definecolor{kit-orange100}{RGB}{223, 155, 27}
\definecolor{kit-orange90}{rgb}{0.8871, 0.6471, 0.1953}
\definecolor{kit-orange80}{rgb}{0.8996, 0.6863, 0.2847}
\definecolor{kit-orange75}{rgb}{0.9059, 0.7059, 0.3294}
\definecolor{kit-orange70}{rgb}{0.9122, 0.7255, 0.3741}
\definecolor{kit-orange60}{rgb}{0.9247, 0.7647, 0.4635}
\definecolor{kit-orange50}{rgb}{0.9373, 0.8039, 0.5529}
\definecolor{kit-orange40}{rgb}{0.9498, 0.8431, 0.6424}
\definecolor{kit-orange30}{rgb}{0.9624, 0.8824, 0.7318}
\definecolor{kit-orange25}{rgb}{0.9686, 0.902, 0.7765}
\definecolor{kit-orange20}{rgb}{0.9749, 0.9216, 0.8212}
\definecolor{kit-orange15}{rgb}{0.9812, 0.9412, 0.8659}
\definecolor{kit-orange10}{rgb}{0.9875, 0.9608, 0.9106}
\definecolor{kit-orange5}{rgb}{0.9937, 0.9804, 0.9553}

\definecolor{kit-lightgreen}{RGB}{140, 182, 60}
\colorlet{KITlightgreen}{kit-lightgreen}
\definecolor{kit-lightgreen100}{RGB}{140, 182, 60}
\definecolor{kit-lightgreen90}{rgb}{0.5941, 0.7424, 0.3118}
\definecolor{kit-lightgreen80}{rgb}{0.6392, 0.771, 0.3882}
\definecolor{kit-lightgreen75}{rgb}{0.6618, 0.7853, 0.4265}
\definecolor{kit-lightgreen70}{rgb}{0.6843, 0.7996, 0.4647}
\definecolor{kit-lightgreen60}{rgb}{0.7294, 0.8282, 0.5412}
\definecolor{kit-lightgreen50}{rgb}{0.7745, 0.8569, 0.6176}
\definecolor{kit-lightgreen40}{rgb}{0.8196, 0.8855, 0.6941}
\definecolor{kit-lightgreen30}{rgb}{0.8647, 0.9141, 0.7706}
\definecolor{kit-lightgreen25}{rgb}{0.8873, 0.9284, 0.8088}
\definecolor{kit-lightgreen20}{rgb}{0.9098, 0.9427, 0.8471}
\definecolor{kit-lightgreen15}{rgb}{0.9324, 0.9571, 0.8853}
\definecolor{kit-lightgreen10}{rgb}{0.9549, 0.9714, 0.9235}
\definecolor{kit-lightgreen5}{rgb}{0.9775, 0.9857, 0.9618}

\definecolor{kit-purple}{RGB}{163, 16, 124}
\colorlet{KITpurple}{kit-purple}
\definecolor{kit-purple100}{RGB}{163, 16, 124}
\definecolor{kit-purple90}{rgb}{0.6753, 0.1565, 0.5376}
\definecolor{kit-purple80}{rgb}{0.7114, 0.2502, 0.589}
\definecolor{kit-purple75}{rgb}{0.7294, 0.2971, 0.6147}
\definecolor{kit-purple70}{rgb}{0.7475, 0.3439, 0.6404}
\definecolor{kit-purple60}{rgb}{0.7835, 0.4376, 0.6918}
\definecolor{kit-purple50}{rgb}{0.8196, 0.5314, 0.7431}
\definecolor{kit-purple40}{rgb}{0.8557, 0.6251, 0.7945}
\definecolor{kit-purple30}{rgb}{0.8918, 0.7188, 0.8459}
\definecolor{kit-purple25}{rgb}{0.9098, 0.7657, 0.8716}
\definecolor{kit-purple20}{rgb}{0.9278, 0.8125, 0.8973}
\definecolor{kit-purple15}{rgb}{0.9459, 0.8594, 0.9229}
\definecolor{kit-purple10}{rgb}{0.9639, 0.9063, 0.9486}
\definecolor{kit-purple5}{rgb}{0.982, 0.9531, 0.9743}

\definecolor{kit-brown}{RGB}{167, 130, 46}
\colorlet{KITbrown}{kit-brown}
\definecolor{kit-brown100}{RGB}{167, 130, 46}
\definecolor{kit-brown90}{rgb}{0.6894, 0.5588, 0.2624}
\definecolor{kit-brown80}{rgb}{0.7239, 0.6078, 0.3443}
\definecolor{kit-brown75}{rgb}{0.7412, 0.6324, 0.3853}
\definecolor{kit-brown70}{rgb}{0.7584, 0.6569, 0.4263}
\definecolor{kit-brown60}{rgb}{0.7929, 0.7059, 0.5082}
\definecolor{kit-brown50}{rgb}{0.8275, 0.7549, 0.5902}
\definecolor{kit-brown40}{rgb}{0.862, 0.8039, 0.6722}
\definecolor{kit-brown30}{rgb}{0.8965, 0.8529, 0.7541}
\definecolor{kit-brown25}{rgb}{0.9137, 0.8775, 0.7951}
\definecolor{kit-brown20}{rgb}{0.931, 0.902, 0.8361}
\definecolor{kit-brown15}{rgb}{0.9482, 0.9265, 0.8771}
\definecolor{kit-brown10}{rgb}{0.9655, 0.951, 0.918}
\definecolor{kit-brown5}{rgb}{0.9827, 0.9755, 0.959}

\definecolor{kit-cyan}{RGB}{35, 161, 224}
\colorlet{KITcyan}{kit-cyan}
\colorlet{KITcyanblue}{kit-cyan}
\definecolor{kit-cyan100}{RGB}{35, 161, 224}
\definecolor{kit-cyan90}{rgb}{0.2235, 0.6682, 0.8906}
\definecolor{kit-cyan80}{rgb}{0.3098, 0.7051, 0.9027}
\definecolor{kit-cyan75}{rgb}{0.3529, 0.7235, 0.9088}
\definecolor{kit-cyan70}{rgb}{0.3961, 0.742, 0.9149}
\definecolor{kit-cyan60}{rgb}{0.4824, 0.7788, 0.9271}
\definecolor{kit-cyan50}{rgb}{0.5686, 0.8157, 0.9392}
\definecolor{kit-cyan40}{rgb}{0.6549, 0.8525, 0.9514}
\definecolor{kit-cyan30}{rgb}{0.7412, 0.8894, 0.9635}
\definecolor{kit-cyan25}{rgb}{0.7843, 0.9078, 0.9696}
\definecolor{kit-cyan20}{rgb}{0.8275, 0.9263, 0.9757}
\definecolor{kit-cyan15}{rgb}{0.8706, 0.9447, 0.9818}
\definecolor{kit-cyan10}{rgb}{0.9137, 0.9631, 0.9878}
\definecolor{kit-cyan5}{rgb}{0.9569, 0.9816, 0.9939}

\definecolor{kit-gray}{RGB}{0, 0, 0}
\colorlet{KITgray}{kit-gray}
\definecolor{kit-gray100}{RGB}{0, 0, 0}
\definecolor{kit-gray90}{rgb}{0.1, 0.1, 0.1}
\definecolor{kit-gray80}{rgb}{0.2, 0.2, 0.2}
\definecolor{kit-gray75}{rgb}{0.25, 0.25, 0.25}
\definecolor{kit-gray70}{rgb}{0.3, 0.3, 0.3}
\definecolor{kit-gray60}{rgb}{0.4, 0.4, 0.4}
\definecolor{kit-gray50}{rgb}{0.5, 0.5, 0.5}
\definecolor{kit-gray40}{rgb}{0.6, 0.6, 0.6}
\definecolor{kit-gray30}{rgb}{0.7, 0.7, 0.7}
\definecolor{kit-gray25}{rgb}{0.75, 0.75, 0.75}
\definecolor{kit-gray20}{rgb}{0.8, 0.8, 0.8}
\definecolor{kit-gray15}{rgb}{0.85, 0.85, 0.85}
\definecolor{kit-gray10}{rgb}{0.9, 0.9, 0.9}
\definecolor{kit-gray5}{rgb}{0.95, 0.95, 0.95}

\definecolor{KITpalegreen}{RGB}{130,190,60}
\colorlet{kit-maigreen100}{KITpalegreen}
\colorlet{kit-maigreen70}{KITpalegreen!70}
\colorlet{kit-maigreen50}{KITpalegreen!50}
\colorlet{kit-maigreen30}{KITpalegreen!30}
\colorlet{kit-maigreen15}{KITpalegreen!15}


\title{State Aware Traffic Generation for Real-Time Network Digital Twins\\

\thanks{This research was supported by the German Federal Ministry of Research, Technology and Space
 (BMFTR) under grant number 16KIS2259 (SUSTAINET-inNOvAte). The responsibility
lies with the authors.
}
}

\author{\IEEEauthorblockN{Enes Koktas}
\IEEEauthorblockA{\textit{Communications Engineering Lab} \\
\textit{Karlsruhe Institute of Technology}\\
Karlsruhe, Germany }
\and
\IEEEauthorblockN{Peter Rost}
\IEEEauthorblockA{\textit{Communications Engineering Lab} \\
\textit{Karlsruhe Institute of Technology}\\
Karlsruhe, Germany} 
}

\maketitle

\begin{abstract}
Digital twins (DTs) enable smarter, self-optimizing mobile networks, but they rely on a steady supply of real world data. Collecting and transferring complete traces in real time is a significant challenge. We present a compact traffic generator that combines hidden Markov model, capturing the broad rhythms of buffering, streaming and idle periods, with a small feed forward mixture density network that generates realistic payload sizes and inter-arrival times to be fed to the DT. This traffic generator trains in seconds on a server GPU, runs in real time and can be fine tuned inside the DT whenever the statistics of the generated data do not match the actual traffic. This enables operators to keep their DT up to date without causing overhead to the operational network. The results show that the traffic generator presented is able to derive realistic packet traces of payload length and inter-arrival time across various metrics that assess distributional fidelity, diversity, and temporal correlation of the synthetic trace.

\end{abstract}


\section{Introduction}
\acp{DT} have recently gained more attention in mobile communications 
because they enable
autonomy, optimization, real-time monitoring, predictive maintenance, fault detection, root cause analysis, and risk-free what-if scenario analysis for current and future communication protocols and implementations. A \ac{DT} framework is composed of three primary elements: the real space, the virtual space, and a link, which functions as a means of communication between the two \cite{9899718}. The \ac{DT} resides in the virtual space and provides a real-time replica of its physical counterpart. \acp{DT} may play an important role for next generation networks because the current network operations often rely on manual modification of system parameters by experts in case of failure or scheduled events. 
There are numerous use cases of \acp{DT} at different layers of a communication system, requiring different granularity of available statistics. Exemplary use cases are resource allocation, handover optimization, traffic offloading, and network planning. 
A network \ac{DT} can integrate data collection, large-scale data processing, and AI modeling to analyze the state of the network as well as to predict future patterns and better coordinate predictive maintenance \cite{irtf-nmrg}.


\subsection{Problem Definition}
One of the most important and challenging aspects of building a \ac{DT} is the data collection process which is key to make sure that the \ac{DT} properly reflects the current network state. The promise of the \ac{DT} is to provide real-time monitoring and optimization. Hence, it must to be fed with data that represents the current behavior of the entities in the real network. Without real-time data tracking and generation, a \ac{DT} would fail for most of the use cases such as scheduling, link adaptation, and link failure prevention.

Traditional network traffic generation techniques rely on idealized statistical models such as Poisson, lognormal or exponential distributions. However, these approaches are not suitable for \ac{DT} operations, because real‐world traffic strongly depends on the actual service.
Moreover, network slicing allows for providing specific quality of service in terms of latency, reliability, and throughput for isolated slices. A \ac{DT} for such a system offering network slicing has to resemble the traffic demand as it is perceived within each slice in order to enable testing of slice-specific algorithms. 

A \ac{DT} applicable to a \ac{RAN} is required to adapt to specific deployment areas such as an industrial factory or a university campus, which typically have different traffic demands and services. Considering the fact that individual \acp{DT} are needed to be trained for these specific areas, it would be impractical to transfer huge amounts of data towards the \ac{DT}.
Consequently, novel traffic generation methodologies encompassing data-driven and statistical techniques are needed to ensure the feasibility of \acp{DT}.  

\subsection{Related Work}
Initial synthetic traffic generation methods, such as TRex, NS-3, and Harpoon \cite{3673792}, relied heavily on simulations and required significant human expertise to be built and set up appropriately. The main limitation of these methods is that they rely on user defined templates and may not capture the intricate intra-flow and intra-packet interactions that characterize real-world traffic \cite{3673792}. 
Numerous other works use \acp{GAN} \cite{9703352}, \cite{3423643}, \cite{9148946}, \cite{RING2019156} which train two competing neural networks against each other to synthesize new network traces. \acp{GAN} provide flexibility in the design of the generator and discriminator structure and the loss function. As traditional \ac{GAN} structures are not sufficient to capture the temporal characteristics of real traffic, researchers have utilized recurrent neural networks in the generator and discriminator.
However, \acp{GAN} are quite challenging to train because of the adversarial training process, which might lead to mode collapse
\cite{3423643}. Moreover, accurate initialization of the generator and discriminator and the loss function design are essential to take full advantage of their power \cite{3664655}.

Researchers have also utilized autoencoder models as a way to assist the generation process and reduce the dimensionality of the original data. In \cite{9320384}, authors introduce a hybrid autoencoder and Wasserstein \ac{GAN} that learns latent representations of internet of things packet size sequences, generates more latent samples with the \ac{GAN}, and decodes them into synthetic bidirectional flows. NeCSTGen is presented in \cite{10000731}, a reproducible variational autoencoder and recurrent neural network based pipeline that clusters packets in latent space via \acp{GMM} and connects them with learned transition matrices to synthesize coherent traffic at packet, flow, and aggregate scales.
While both articles reproduce generic traffic traces with high precision, the implemented recurrent layers are rather resource intensive and implicitly include timing behavior in large hidden state vectors. In a \ac{DT} context where each individual \ac{RAN} scenario should be re-trained and executed on site, this translates into longer optimization cycles and higher memory footprints.

In contrast, a compact \ac{HMM} workflow with a small number of states, as we use it in our work, provides an explicit transition structure, converges quickly on small datasets, and can be sampled or re-estimated in real time on inexpensive hardware, making it a more practical backbone for lightweight traffic generation. In \cite{9145646}, the authors synthesized flows with two independent \acp{HMM}, one for payload length and one for \ac{IAT} via the Baum-Welch algorithm. The generators reproduce the marginal length and timing distributions observed in live video streaming and online gaming traces. Treating the features in isolation, however, might miss the joint size–timing dynamics that drive burstiness, queue occupancy, and delay. We therefore propose a single multi-state \ac{HMM} trained on the two dimensional sequence of payload length and \ac{IAT}. This unified model captures the cross correlations and is expected to yield more realistic queuing behavior.


\subsection{Contribution}
In this article, we propose a compact hybrid \ac{HMM} and \ac{MDN} \cite{astonpr373} pipeline that explicitly factorizes a flow into (i) a small, interpretable state backbone, which captures coarse behavioral modes such as \emph{buffering}, \emph{steady streaming} and \emph{idle} and (ii) a \ac{MDN} that learns, for each state, a mixture of Gaussians over the fundamental packet tuple of payload length and \ac{IAT}. The \ac{HMM}’s transition matrix is estimated via a single \ac{EM} iteration and comprises only a few dozen parameters \cite{bilmes97}. The feed forward \ac{MDN}, which is conditioned on one-hot state vectors, compresses the high dimensional packet space into a set of mixture weights, means and variances that can be sampled in real time. This modular design lets a \ac{DT} re-estimate the \ac{HMM} or fine-tune the \ac{MDN} on the fly when network conditions evolve, providing a real-time traffic model required by \ac{DT} applications.


\section{Dataset and notation}
\begin{figure}[t]
    \centerline{\includegraphics[trim={0.5cm 0.5cm 0.5cm 0.5cm},clip,width=7.5cm]{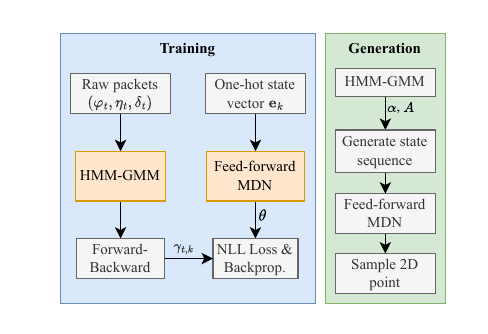}}
    \caption{Training and generation pipeline of the proposed traffic generator.}
    \label{fig:pipeline_arch}
\end{figure}

The proposed traffic generator splits the packet level dynamics of an application flow into two cooperating parts: a compact hidden state process and a flexible neural emission model. The network traffic can be thought of as moving through different states (e.g. buffering, steady streaming, idle) depending on the application type. These state transitions can be modeled with an \ac{HMM}, while the emission distribution of each state is represented by \ac{MDN}. This leads to a  hybrid approach in which an \ac{HMM} supplies a state sequence, and an \ac{MDN} refines each state with a rich mixture of Gaussians. 
Schematic overviews of the training and generation phases are shown in
Fig.~\ref{fig:pipeline_arch}. There are multiple types of \acp{HMM}.  We are using an \ac{HMM} with \ac{GMM} emissions, known as \ac{HMM}-\ac{GMM} \cite{18626}. It is important to have \ac{GMM} emissions in \ac{HMM} since they are used to calculate the state posteriors. 

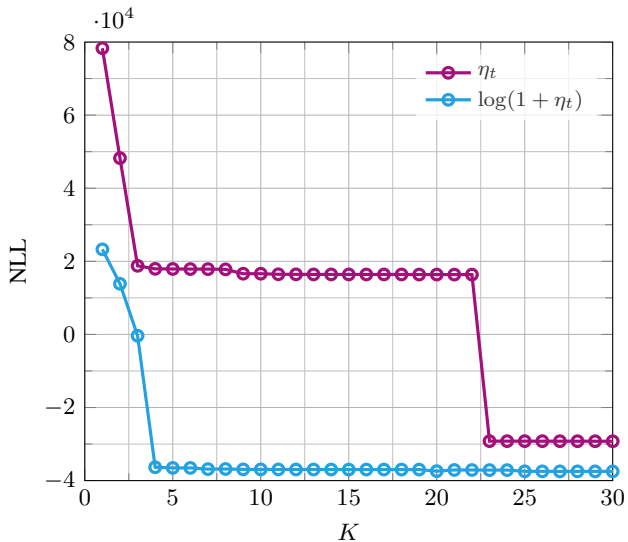
\begin{figure}[t]
    \centering
    \begin{tikzpicture}
        \centering
        \begin{axis}[
            grid style={line width=.2pt, gray!50},
            major grid style={line width=.3pt, gray!60},
            minor tick num=1,
            legend style={font=\small, draw=none, fill=white, fill opacity=.8},
            tick label style={font=\small},
            ylabel style={font=\small},
            xlabel style={font=\small},
            width=8.6cm,
            xlabel={$K$},
            ylabel={NLL},
            xmin=0, xmax=30,
            ymin=-40000, ymax=80000,
            grid=both,
            legend cell align={left},
            legend pos=north east,
            legend style={font=\footnotesize},
        ]
        
        \addplot[kit-purple100, mark=o, very thick] table [x index=0, y index=1, col sep=comma] {data/nll_comp_df_raw_HTTP.csv};
        \addlegendentry{$\eta_t$}
        \addplot[kit-cyan100, mark=o, very thick] table [x index=0, y index=2, col sep=comma] {data/nll_comp_df_raw_HTTP.csv};
        \addlegendentry{$\log(1+\eta_t)$}
        \end{axis}
    \end{tikzpicture}
    \vspace{-0.2cm}
    \caption{\ac{NLL} of a plain \ac{GMM} fitted on $\eta_t$ vs $K$ for HTTP traffic}
    \label{fig:log_transform}
\end{figure}

We use two real public packet traces, which have already been processed including flow identification and labeling\footnote{https://github.com/fmeslet/NeCSTGen}, one comprise of \ac{HTTP} and another one by user \ac{UDP} traffic. 
The notation of a packet in the dataset is described by the triplet $\mathcal{D} = [(\varphi_t,\eta_t,\delta_t)]_{t=1}^{T}$, the sequence of $T$ packets, where $\varphi_t\in\mathbb{N}$ is the flow identifier, $\eta_t\in\mathbb R{\ge0}$ is the payload length in bytes, and $\delta_t\in
\mathbb R{\ge0}$ is the \ac{IAT} measured in seconds between the current and the previous
packet. Basic cleaning removes non-finite samples and the rare, heavy-tailed packets with $\delta_t > 25$ ms (i.e.\ those above the $98$th percentile of the trace), because such outliers destabilize the HMM posteriors and the feed-forward MDN.
Packet sizes also follow a heavily tailed probability density function dominated by nearly \ac{MTU} sized segments forming a distant peak and very short acknowledgment messages. The \ac{GMM} of the \ac{HMM} must use many components just to stretch across that tail. In contrast, applying the natural log to $\eta_t$ concentrates the bulk near the center, and achieves the same \ac{NLL} with roughly one-quarter of the components as shown in Fig. ~\ref{fig:log_transform}. In practice, this gives the \ac{HMM} cleaner state boundaries. 
The raw feature vector is defined as
\begin{equation}
\bm{x}_t = [\log(1+\eta_t),\,\delta_t]^{\intercal} \in \mathbb{R}^{2},
\end{equation}
where $(\cdot)^\intercal$ denotes the transpose operator. 
The normalization is then applied as 
\begin{equation}
    \bm{z}_t = \frac{\bm{x}_t-\bm{\mu}}{\bm{\sigma}}.
\end{equation}
The empirical means and standard deviations
are computed over the entire training set so that the transformation is shared across all flows. The dataset is then split by \emph{flow identifier} so that the temporal coherence of
individual sessions is preserved. $90\%$ of the flows form the training set and the remaining
$10\%$ constitute the test set.

\section{System Description}\label{sec:system_model}
\subsection{The HMM-GMM}
We assume that a packet sequence alternates between $K$ coarse behavioral modes such as buffering,
steady streaming, or idle waiting. These modes are represented by the discrete hidden state
process $\{s_t\}_{t=1}^{T}$ with $s_t\in\{1,\dots,K\}$. Transitions follow a first–order Markov
chain with initial distribution $\alpha_k=\Pr(s_{1}=k)$ and transition matrix
$A$, whose element $A_{ij}=\Pr(s_{t}=j\mid s_{t-1}=i)$. Conditioned on the
state, the normalized observation $\bm{\hat{z}}_t$ is drawn from the \ac{GMM} with $J$ components as
\begin{equation} \label{eq:gmm}
\begin{aligned}
p(\bm{\hat{z}}_t\mid s_t=k) &=\sum_{j=1}^{J} \beta_{k,j}\,\times
  \\
&\quad\mathcal{N}\left(\bm{z}_t\middle\vert\bm{\mu}_{k,j}^{[0]},\operatorname{diag}\left(\left(\bm{\sigma}_{k,j}^{[0]}\right)^{2}\right)\right),
\end{aligned}
\end{equation}
where $\mathcal{N}\left(z\mid \mu, \Sigma\right)$ denotes the multivariate normal density evaluated at $z$ with mean vector $\mu$ and covariance matrix $\Sigma$ and $\operatorname{diag}(v)$ denotes the diagonal matrix whose entries on the main diagonal are given by the elements of the vector $v$. The number of states $K$ and the number of Gaussians per state $J$ are hyper‑parameters and $\beta_{k,j}\geq0$ are the mixture weights, with $\sum_j \beta_{k,j} = 1$, learned by the \ac{EM} step. $K$ and $J$ are selected separately for each application type by an automated Weights and Biases sweep \cite{weights_&_biases_2017}. All remaining \ac{HMM}-\ac{GMM} parameters $\bm{\Theta}=\left(\bm{\alpha},\bm{A},\bm{\beta},\bm{\mu}^{[0]},\bm{\sigma}^{[0]}\right)$
are obtained by running the \ac{EM} algorithm on all training flows. Superscript $(\cdot)^{[0]}$ denotes the \ac{GMM} parameters in the \ac{HMM} and $(\cdot)^{[1]}$ denotes Gaussian parameters in the \ac{MDN}. At convergence the forward–backward algorithm provides the state posteriors as
\begin{equation}
\gamma_{t,k}=p_{\bm{\Theta}}(s_t=k\mid \bm{z}_{1:T}).
\label{eq:gamma}
\end{equation}
These posteriors will serve as importance weights for the supervised training of the \ac{MDN}.

\subsection{The feed forward MDN}
The \ac{HMM}-\ac{GMM} is sufficient to bootstrap the hidden structure, yet it is unable to capture the multimodality exhibited by real packet distributions. Therefore, the parametric emission of each state is replaced with a conditional mixture learned by a feed forward \ac{MDN}. The \ac{HMM}-\ac{GMM} yields the state posteriors, which will be used to perform weighted backpropagation of the \ac{NLL} during the training of the \ac{MDN}. The state sequence sampled from the \ac{HMM}-\ac{GMM}, on the other hand, will be used to gather conditional samples from the \ac{MDN} during generation.
\ac{MDN} has two layers and $128$ tanh units per layer. For a one-hot state encoding $\bm{e}_k$, the feed forward \ac{MDN} $f_{\bm{\theta}}\colon \mathbb{R}^{K}\to \mathbb{R}^{M\times(1+2+2)}$ outputs, for each mixture component $m=1,\dots,M$, a mixture weight $\pi_{k,m}$, a two-dimensional mean $\bm{\mu}_{k,m}^{[1]}$, and a two-dimensional variance $\left(\bm{\sigma}_{k,m}^{[1]}\right)^{2}$. The weights $\pi_{k,m}$ are normalized via a softmax to form valid probabilities, and each variance is passed through a softplus to guarantee positivity.
The conditional density becomes
\begin{equation}
\begin{aligned}
p_{\bm{\theta}}(\bm{z}_t\mid s_t=k) &=\sum_{m=1}^{M}\pi_{k,m}\,\times \\
& \quad
\mathcal{N}\left(
\bm{z}_t\mid \bm{\mu}_{k,m}^{[1]},\operatorname{diag}\left(\left(\bm{\sigma}_{k,m}^{[1]}\right)^{2}\right)\right).
\label{eq:mdn_density}
\end{aligned}
\end{equation}
The \ac{GMM} and the \ac{MDN} given in ~\eqref{eq:gmm} and ~\eqref{eq:mdn_density}, respectively, might seem identical because both of them are Gaussian mixtures. However, the way they are obtained and the purpose they serve are different. The \ac{GMM} is estimated once by the \ac{EM} step directly from data while the \ac{MDN} is produced by a two-layer feed forward network. The \ac{GMM} provides the state posteriors $\gamma_{t,k}$, expressing the probability of a packet belonging to a particular state. On the other hand, the \ac{MDN} uses the posteriors as soft targets and learns to output a richer mixture that refines the coarse shape from the \ac{HMM}.

\begin{algorithm}[b]
\caption{Train Hybrid HMM--MDN generator}\label{alg:train}
\begin{algorithmic}[1]
\Require $\mathcal{D}=\{\bm{z}_t\}_{t=1}^T$, $K$, $J$, $M$
\Ensure  $\bm{\Theta}$, $\bm{\theta}$
\State \textbf{// Learn state dynamics} 
\State Fit a $K$-state, $J$-component HMM-\ac{GMM} on $\mathcal{D}$ via EM $\;\rightarrow\;$ $\bm{\Theta}$
\State Compute posteriors $\gamma_{t,k}=p_{\bm{\Theta}}(s_t=k\mid \bm{z}_{1:T})$ for all $t,k$
\State \textbf{// Refine state conditioned emissions}
\State Build weighted set $\mathcal{B}=\bigl\{\bigl(\bm{e}_k,\bm{z}_t,w_{t,k}\bigr):w_{t,k}=\gamma_{t,k}\bigr\}$
\For{$\text{epoch}=1\to E$}
    \For{\textbf{minibatch} $(\bm{e}_k,\bm{z},\bm{w})\subset\mathcal{B}$}
        \State $(\bm{\pi},\bm{\mu}^{[1]},\bm{\sigma}^{[1]})\gets\text{MDN}_{\bm{\theta}}(\bm{e}_k)$
        \State $\ell\gets \sum w\,\bigl[-\!\log p_{\bm{\theta}}(\bm{z}\mid s=k)\bigr]$   
        \State $\bm{\theta} \leftarrow \bm{\theta}-\eta\nabla_{\bm{\theta}}\ell$ \Comment{Adam, grad-clip $5$}
    \EndFor
\EndFor
\State \Return $\bm{\Theta},\;\bm{\theta}$
\end{algorithmic}
\end{algorithm}

\subsection{Weighted maximum likelihood training}
Training proceeds in two stages executed once per application protocol. First, the \ac{HMM}-\ac{GMM} parameters
are fitted on the training flows as explained above, yielding the posterior weights
$\gamma_{t,k}$. Secondly, the \ac{MDN} parameters $\bm{\theta}$ are optimized by minimizing the
weighted \ac{NLL} of ~\eqref{eq:mdn_density}:
\begin{align} \label{eq:nll}
\mathcal{L}(\bm{\theta}) &= \frac{1}{\sum_{t,k}w_{t,k}}\sum_{t,k}w_{t,k}\,\ell_{t,k}(\bm{\theta}),
\end{align}
where 
\begin{equation}
\ell_{t,k}(\bm{\theta}) = -\log p_{\bm{\theta}}(\bm{z}_{t,k}\mid s_{t,k})
\end{equation} 
and $w_{t,k} = |B|\,\gamma_{t,k}$, where $|B|$ is the number of rows in the current mini-batch. The variable $w_{t,k}$ rescales the posteriors so that their sum equals $|B|$. Hence, the denominator in ~\eqref{eq:nll} normalizes the batch back to an average per-sample loss, which stabilizes the optimizer's learning rate. Thanks to the weighting of the \ac{NLL} with the posteriors, packets with $\gamma \approx 1$ for state $k$ train only the $k$-th row of the \ac{MDN}, letting it specialize and push the $M$ components where they are really needed. Without those weights, the \ac{MDN} would see a scrambled mix of modes and learn an average of the available features. Lastly, gradients are backpropagated using the Adam optimizer with a learning rate of $1\times 10^{-3}$ and clipped to an Euclidean norm of $5$ to avoid exploding updates. The Pseudo-code of the training phase is given in Algorithm~\ref{alg:train}.

\subsection{Generation and Complexity}
During generation, the \ac{HMM}-\ac{GMM} generates a state trajectory of the desired length. Starting from a state sampled from $\bm{\alpha}$ we draw subsequent states recursively from the transition matrix $\bm{A}$. For each state $k$, the \ac{MDN} outputs are evaluated once and cached. Sampling a packet proceeds in two steps: (i) select a mixture component index $m$ according to the categorical distribution $\mathrm{Cat}(\bm{\pi}_{s_t})$, and (ii) draw $\bm{\hat{z}}_t$ from the Gaussian of that component \cite{astonpr373}. 
The \ac{MDN} generates samples in the normalized space. They are converted back to physical units by applying the inverse transform with the standard deviation and the mean recorded during the initial normalization. For $\eta_t$, also the previous log-transform is reversed. The resulting $\eta_t$ is clipped to the \ac{MTU} and rounded to the nearest byte. Non-positive $\delta_t$ are replaced by $\unit[1]{\mu s}$. The Pseudo-code of the generation phase is given in Algorithm \ref{alg:gen}.

\begin{algorithm}[b]
\caption{Generate synthetic flow of length $L$}\label{alg:gen}
\begin{algorithmic}[1]
\Require $L$, $\bm{\Theta}$, $\bm{\theta}$
\Ensure  flow $\{(\hat{\eta}_t,\hat{\delta}_t)\}_{t=1}^{L}$
\State Pre-compute $(\bm{\pi}_k,\bm{\mu}^{[1]}_k,\bm{\sigma}^{[1]}_k)=\text{MDN}_{\bm{\theta}}(\bm{e}_k)$ for $k=1,\dots,K$
\State $s_1\sim\text{Cat}(\bm{\alpha})$
\For{$t=1\;\textbf{to}\;L$}
    \If{$t>1$} 
    \State $\bm{p} \gets \bm{A}_{s_{t-1}}$   
\State $s_t \sim \mathrm{Cat}(\bm{p})$ \EndIf
    \State $m \sim \mathrm{Cat}(\bm{\pi}_{s_t})$
    \State $\bm{\hat{z}}_t\sim\mathcal{N}\!\left(\bm{\mu}^{[1]}_{s_t,m},\;\mathrm{diag}\left(\left(\bm{\sigma}^{[1]}_{s_t,m}\right)^{2}\right)\right)$
    \State $(\hat{\eta}_t,\hat{\delta}_t)\gets\text{inverse\_transform}(\bm{\hat{z}}_t)$
\EndFor
\State \Return $\{(\hat{\eta}_t,\hat{\delta}_t)\}_{t=1}^{L}$
\end{algorithmic}
\end{algorithm}


The model stores $K^{2}+K$ transition parameters for the \ac{HMM}-\ac{GMM} and,
$Q(K + Q + 5M) + 2Q + 5M$ scalar weights for the feed forward \ac{MDN}, where $Q$ is the hidden layer width.  
With $K = 6$, $M = 32$, and two hidden layers of $Q = 128$, this results in approximately $3.8 \times 10^{4}$ floating point numbers. 
Because of the explicit separation between state dynamics and emissions, the \ac{HMM}-\ac{GMM} can be re–estimated on a short sliding window of recent traffic while the \ac{MDN} remains fixed, enabling rapid adaptation to changing network conditions.

\section{Evaluation and Results}
Evaluation of the similarity between real and synthetic traffic is assessed from multiple perspectives such as distributional fidelity, diversity, and temporal correlation. 
Methods such as the \ac{KL} divergence, \ac{KS} test, \ac{ACF}, and \ac{PSD} are utilized to measure the realism of synthetically generated data \cite{schoen}.

\subsection{Metrics}
The effectiveness of the proposed algorithm is measured with twelve complementary statistics that look at the traffic from three perspectives; distributional fidelity, diversity, and temporal correlation. 
The metrics associated with distributional fidelity are given as
$$
\begin{aligned}
D_{\mathrm{KS}}^{y}   &= \sup_{u}\bigl|\bar{F}_{y}(u)-\bar{F}_{\hat{y}}(u)\bigr|,\\
D_{\mathrm{KL}}^{\text{CDF},y}   &= D_{\mathrm{KL}}\!\bigl(\partial_u\bar{F}_{y},\partial_u\bar{F}_{\hat{y}}\bigr),
\end{aligned}
$$
where $y\in\{\eta_t,\delta_t\}$, $\bar{F}_{y}$ is the average empirical \ac{CDF} taken over all flows, $\sup\limits_{u}\bigl|\cdot\bigr|$ is the supremum over all possible threshold values $u$, and $\partial_u$ is the partial derivative with respect to $u$. The metrics associated with temporal correlation are given as
$$
\begin{aligned}
D_{\mathrm{KL}}^{\text{PSD},y}   &= D_{\mathrm{KL}}\!\bigl(p_{y},p_{\hat{y}}\bigr),\\
& = \sum\limits_y p_{y}(y)\cdot \log\left(\frac{p_{y}(y)}{p_{\hat y}(y)}\right) \\
\mathrm{ACF}_{\mathrm{RMSE}}^{y}   &= \sqrt{\tfrac1{L}\sum_{\ell=1}^{L}\!\bigl(r_{y}[\ell]-r_{\hat{y}}[\ell]\bigr)^{2}},
\end{aligned}
$$
where $p_{y}(\cdot)$ is the \ac{PSD} of $y$ determined over all flows and $r_{y}[\ell]$ is the lag-$\ell$ autocorrelation. The metrics associated with diversity are given as
$$
\begin{aligned}
\Delta\mu_{H}^{y}   &= \bigl|\mathbb{E}[H_{y}^{(i)}]-\mathbb{E}[
H_{\hat{y}}^{(i)}]\bigr|,\\
C_{H}^{y}           &= \frac{\max\{0,H_{\hat{y}}^{\max}-H_{y}^{\max}\}+\max\{0,H_{y}^{\min}-H_{\hat{y}}^{\min}\}}
{H_{y}^{\max}-H_{y}^{\min}},
\end{aligned}
$$
where $H_{y}^{(i)}$ is the spectral entropy of the $i$-th flow, $H_{y}^{\max}$ and $H_{y}^{\min}$ are the maximum and minimum entropies observed across flows. Every metric is a non-negative divergence, therefore lower values are preferable.
Summing up all these metrics lets Weights \& Biases sweeps rank candidates automatically. The search initialized with the \ac{HMM}-\ac{GMM}. We iterate over the number of \ac{HMM} states, mixture components, variance floor \emph{min\_covar} and \ac{EM} limit \emph{max\_iter} as shown in Table~\ref{tab:metrics}. The \ac{HMM}-\ac{GMM} with the lowest aggregate score yields state posteriors $\gamma_{t,k}$, which are then fixed and reused as soft weights when optimizing the \ac{MDN} parameters. This two stage optimization scheme allows us to re-tune either stage in isolation if the required realism is no longer sustained. Table~\ref{tab:parameters2} lists the optimal values evaluated for \ac{HTTP} and \ac{UDP} traffic. Table~\ref{tab:metrics} shows the metrics calculated with those parameters and indicates which perspective each metric addresses.

{
\renewcommand{\arraystretch}{1.4}
\begin{table}[t!]
\centering
\caption{Optimized parameters of HTTP and UDP traffic generators}
\label{tab:parameters2}
\resizebox{\columnwidth}{!}{%
\begin{tabular}{|l|lllllll|}
\hline
\multirow{3}{*}{\textbf{Protocol}} &
  \multicolumn{7}{c|}{\textbf{Parameter}} \\ \cline{2-8} 
 &
  \multicolumn{4}{c|}{\textit{\textbf{\ac{HMM}-\ac{GMM}}}} &
  \multicolumn{3}{c|}{\textit{\textbf{\ac{MDN}}}} \\ \cline{2-8} 
 &
  \multicolumn{1}{l|}{$K$} &
  \multicolumn{1}{l|}{$J$} &
  \multicolumn{1}{l|}{min\_covar} &
  \multicolumn{1}{l|}{max\_iter} &
  \multicolumn{1}{l|}{$H$} &
  \multicolumn{1}{l|}{$M$} &
  num\_epochs \\ \hline
\ac{HTTP} &
  \multicolumn{1}{l|}{$3$} &
  \multicolumn{1}{l|}{$3$} &
  \multicolumn{1}{l|}{$10^{-3}$} &
  \multicolumn{1}{l|}{$200$} &
  \multicolumn{1}{l|}{$128$} &
  \multicolumn{1}{l|}{$12$} &
  $90$ \\ \hline
\ac{UDP} &
  \multicolumn{1}{l|}{$6$} &
  \multicolumn{1}{l|}{$7$} &
  \multicolumn{1}{l|}{$10^{-2}$} &
  \multicolumn{1}{l|}{$200$} &
  \multicolumn{1}{l|}{$128$} &
  \multicolumn{1}{l|}{$32$} &
  $30$ \\ \hline
\end{tabular}%
}
\end{table}
}

\subsection{Results}
All primitive scores remain small. For \ac{HTTP}, the \ac{KS} gaps are $0.21$ for $\eta$ and $0.09$ for $\delta$, both lower than the $0.36/0.37$ we obtained from our re-implementation of \cite{9320384} and the $0.65/0.11$ computed from the published samples of \cite{10000731}. Additionally, all KS scores lie below the $0.38$–$0.85$ range reported for four GAN variants in \cite{Anande03042023}.
The \ac{PSD} \ac{KL} divergences stay below $0.12$, the \ac{CDF} divergences are $6.53$ and $3.10$, and the entropy-bias and diversity penalties are each under $0.28$. 
Together, these results show that the model matches first-order marginals, captures intra-flow correlations in both time and frequency domains, and preserves the diversity of the flows.

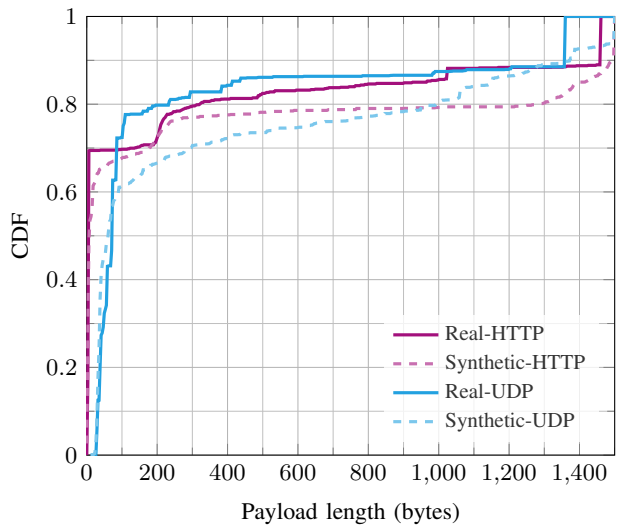
\begin{figure}[t]
    \centering
    \begin{tikzpicture}
        \begin{axis}[
            grid style={line width=.2pt, gray!50},
            major grid style={line width=.3pt, gray!60},
            minor tick num=1,
            legend style={font=\small, draw=none, fill=white, fill opacity=.8},
            tick label style={font=\small},
            ylabel style={font=\small},
            xlabel style={font=\small},
            width=8.6cm,
            xlabel={Payload length (bytes)},
            ylabel={CDF},
            xmin=0, xmax=1500,
            ymin=0, ymax=1,
            grid=both,
            legend cell align={left},
            legend pos=south east,
            legend style={font=\footnotesize},
        ]
        
        \addplot[kit-purple100, solid, very thick] table [x index=0, y index=1, col sep=comma] {data/payload_cdf_http.csv};
        \addlegendentry{Real-HTTP}
        \addplot[kit-purple60, dashed, very thick] table [x index=0, y index=2, col sep=comma] {data/payload_cdf_http.csv};
        \addlegendentry{Synthetic-HTTP}
        \addplot[kit-cyan100,  solid,  very thick] table [x index=0, y index=1, col sep=comma] {data/payload_cdf_udp.csv};
        \addlegendentry{Real-UDP}
        \addplot[kit-cyan60, dashed, very thick] table [x index=0, y index=2, col sep=comma] {data/payload_cdf_udp.csv};
        \addlegendentry{Synthetic-UDP}
        \end{axis}
    \end{tikzpicture}
    \caption{Average per-flow \ac{CDF} comparison of payload length for \ac{HTTP} and \ac{UDP} traffic.}
    \label{fig:cdf_payload}
\end{figure}

\begin{figure}[t]
    \centering
    \begin{tikzpicture}
        \begin{axis}[
            grid style={line width=.2pt, gray!50},
            major grid style={line width=.3pt, gray!60},
            minor tick num=1,
            scaled x ticks=false,                 
              xticklabel style={
                /pgf/number format/fixed,           
                /pgf/number format/precision=3      
              },
            legend style={font=\small, draw=none, fill=white, fill opacity=.8},
            tick label style={font=\small},
            ylabel style={font=\small},
            xlabel style={font=\small},
            width=8.6cm,
            xlabel={IAT (seconds)},
            ylabel={CDF},
            xmin=1e-6, xmax=25e-3,
            ymin=0, ymax=1,
            grid=both,
            legend cell align={left},
            legend pos=south east,
            legend style={font=\footnotesize},
        ]
        
        \addplot[kit-purple100, solid, very thick] table [x index=0, y index=1, col sep=comma] {data/time_cdf_http.csv};
        \addlegendentry{Real-HTTP}
        \addplot[kit-purple60, dashed, very thick] table [x index=0, y index=2, col sep=comma] {data/time_cdf_http.csv};
        \addlegendentry{Synthetic-HTTP}
        \addplot[kit-cyan100,  solid,  very thick] table [x index=0, y index=1, col sep=comma] {data/time_cdf_udp.csv};
        \addlegendentry{Real-UDP}
        \addplot[kit-cyan60, dashed, very thick] table [x index=0, y index=2, col sep=comma] {data/time_cdf_udp.csv};
        \addlegendentry{Synthetic-UDP}
        \end{axis}
    \end{tikzpicture}
    \caption{Average per-flow \ac{CDF} comparison of \ac{IAT} for \ac{HTTP} and \ac{UDP} traffic.}
    \label{fig:cdf_iat}
\end{figure}
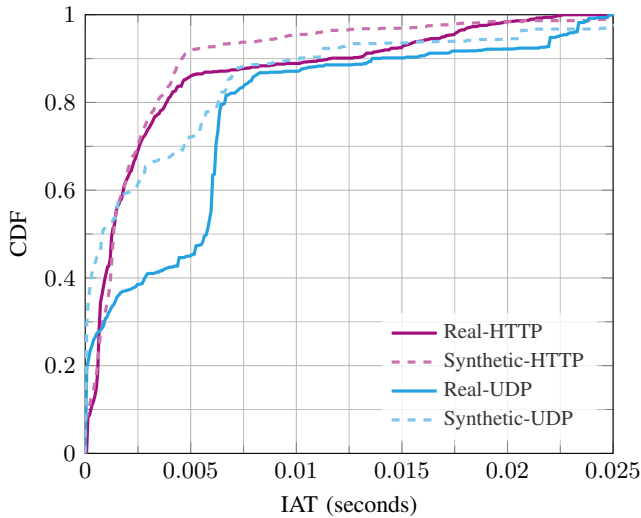

{
\renewcommand{\arraystretch}{1.7}
\begin{table}[t]
    \centering
    \caption{Results of multiple evaluation metrics for synthetic HTTP and UDP traffic.}
    \label{tab:metrics}
    \resizebox{\columnwidth}{!}{%
    \begin{tabular}{|ll|c|c|c|c|c|}
        \hline
        \multicolumn{2}{|c|}{\textbf{Metric}} &
          \textbf{\ac{HTTP}} &
          \textbf{\ac{UDP}} &
          \textbf{\begin{tabular}[c]{@{}c@{}}Distributional \vspace{-0.1cm}\\ Fidelity\end{tabular}} &
          \textbf{Diversity} &
          \textbf{\begin{tabular}[c]{@{}c@{}}Temporal\vspace{-0.1cm}\\ Correlation\end{tabular}} \\ \hline
        \multicolumn{1}{|l|}{\multirow{3}{*}{$D_{\mathrm{KS}}^{\eta}$}} & Ours & $0.2105$ & $0.1407$ & \multirow{3}{*}{\checkmark} & \multirow{3}{*}{} & \multirow{3}{*}{} \\ \cline{2-4}
        \multicolumn{1}{|l|}{} & \cite{9320384} & \multicolumn{1}{l|}{$0.3555$} & \multicolumn{1}{l|}{$0.8553$} &  &  &  \\ \cline{2-4}
        \multicolumn{1}{|l|}{} & \cite{10000731} & \multicolumn{1}{l|}{$0.6466$} & \multicolumn{1}{l|}{$0.2300$} &  &  &  \\ \hline
        \multicolumn{1}{|l|}{\multirow{3}{*}{$D_{\mathrm{KS}}^{\delta}$}} & Ours & $0.0892$ & $0.1962$ & \multirow{3}{*}{\checkmark} & \multirow{3}{*}{} & \multirow{3}{*}{} \\ \cline{2-4}
        \multicolumn{1}{|l|}{} & \cite{9320384} & \multicolumn{1}{l|}{$0.3721$} & \multicolumn{1}{l|}{$0.9111$} &  &  &  \\ \cline{2-4}
        \multicolumn{1}{|l|}{} & \cite{10000731} & \multicolumn{1}{l|}{$0.1063$} & \multicolumn{1}{l|}{$0.3383$} &  &  &  \\ \hline
        \multicolumn{2}{|l|}{$D_{\mathrm{KL}}^{\text{PSD},\eta}$} & $0.1107$ & $0.2134$ &  &  & \checkmark \\ \hline
        \multicolumn{2}{|l|}{$D_{\mathrm{KL}}^{\text{PSD},\delta}$} & $0.0321$ & $0.0251$ &  &  & \checkmark \\ \hline
        \multicolumn{2}{|l|}{$D_{\mathrm{KL}}^{\text{CDF},\eta}$} & $6.5280$ & $8.5130$ & \checkmark &  &  \\ \hline
        \multicolumn{2}{|l|}{$D_{\mathrm{KL}}^{\text{CDF},\delta}$} & $3.1030$ & $4.0230$ & \checkmark &  &  \\ \hline
        \multicolumn{2}{|l|}{$C_{H}^{\eta}$} & $0.2754$ & $0.6585$ &  & \checkmark &  \\ \hline
        \multicolumn{2}{|l|}{$C_{H}^{\delta}$} & $0.2707$ & $0.7092$ &  & \checkmark &  \\ \hline
        \multicolumn{2}{|l|}{$\Delta\mu_{H}^{\eta}$} & $0.1288$ & $0.4420$ &  & \checkmark &  \\ \hline
        \multicolumn{2}{|l|}{$\Delta\mu_{H}^{\delta}$} & $0.1027$ & $0.1805$ &  & \checkmark &  \\ \hline
        \multicolumn{2}{|l|}{$\mathrm{ACF}_{\mathrm{RMSE}}^{\eta}$} & $0.1471$ & $0.1121$ &  &  & \checkmark \\ \hline
        \multicolumn{2}{|l|}{$\mathrm{ACF}_{\mathrm{RMSE}}^{\delta}$} & $0.0067$ & $0.0513$ &  &  & \checkmark \\ \hline

    \end{tabular}%
    }
\end{table}
}
Average per-flow \ac{CDF} comparison of payload length and \ac{IAT} for \ac{HTTP} and \ac{UDP} traffic are shown in Fig.~\ref{fig:cdf_payload} and Fig.~\ref{fig:cdf_iat}, respectively. 
Across almost the entire range the synthetic curves closely match their real counterparts, confirming that the model reproduces both the heavy left tail of very small payloads and the stepped structure that reflects the common \ac{MTU} levels. A minor deviation is visible for \ac{UDP} in Fig. \ref{fig:cdf_iat} at $\delta_{t}\!\approx\!2.5\times10^{-3}\,\text{and} \,6.5\times10^{-3}\,\text{s}$. That deviation corresponds to flows containing only one to three packets, producing an abrupt rise in the empirical \ac{CDF}. 
Their impact on higher-order metrics is negligible because such micro-flows carry less than $5$ \% of the total traffic. Nevertheless we retain them to preserve the true flow-size distribution. 
Overall, the figures show that the generator captures first-order marginals for both $\eta_{t}$ and $\delta_{t}$ while also handling the heterogeneous mixture of short and long flows that characterizes real network traffic.
Training and synthetic generation completes in the range of $20-25$ s for \ac{HTTP} and \ac{UDP} on a single AMD EPYC 7742 node equipped with four NVIDIA RTX A6000 cards. Once trained, sampling runs in real time on an embedded CPU because only the small \ac{HMM} is queried at every packet, while the heavier \ac{MDN} can be pre-evaluated or off-loaded. This satisfies the strict latency budgets of \ac{DT} traffic injection without compromising statistical realism.

\section{Conclusion}
In this article, we introduced a compact hybrid traffic generator in which a \ac{HMM}-\ac{GMM} captures the coarse rhythm of buffering, streaming and idle periods, while a feed forward \ac{MDN} refines each state with a rich Gaussian mixture over payload length and \ac{IAT}. Experiments on real \ac{HTTP} and \ac{UDP} traces showed that this combination reproduces first-order marginals, frequency-domain structure and entropy diversity with promising divergences. The generator trains in under half a minute and runs in real time on an embedded CPU. Such efficiency, together with the transparent transition matrix and the ability to re-estimate the \ac{HMM} on a sliding window, makes the model an ideal building block for network digital twins that must adapt quickly to local traffic while preserving interpretable control over bursts and idle gaps. Although the present generator already mirrors realistic traffic traces, future work will extend the \ac{MDN} to condition its emissions on the residual flow length, so that synthetic traces also respect the empirical distribution of flow sizes and further tighten the link between the virtual and the physical network twin.

\bibliographystyle{IEEEtran}
\bibliography{ref}

\end{document}